\documentclass[preprint,12pt]{elsarticle}

\usepackage{amssymb}
\usepackage{amsmath}
\usepackage{algorithm}      
\usepackage{algpseudocode}  
\usepackage{url}

\usepackage{subfig} 
\journal{Acta Astronautica}

\begin{document}

\begin{frontmatter}

\title{Global Optimization of Multi-Flyby Trajectories for Multi-Orbital-Plane Constellations Inspection}


\author[1]{An-yi Huang}
\author[1]{Hong-xin Shen}
\author[1]{Zhao Li}
\author[1]{Cong Sun}
\author[1]{Chao Sheng}
\author[1]{Zheng-zhong Kuai}

\affiliation[1]{organization={Advanced Orbit Control Group},
            city={Xi'an},
            postcode={710000}, 
            country={China}}

\begin{abstract}
The rapid expansion of mega-constellations in low Earth orbits has posed significant challenges to space traffic management, necessitating periodic inspections of satellites to ensure the sustainability of the space environment when economically feasible. This study addresses the orbital design challenge associated with inspecting numerous satellites distributed across multiple orbital planes through flybys by proposing an innovative orbital-plane-based inspection strategy. The proposed methodology reformulates the multi-satellite flyby problem into a multi-rendezvous trajectory planning problem by proposing an analytical approach to determine a maneuver-free inspection orbit that enables flyby of all satellites within a specific orbital plane. Additionally, a three-layer global optimization framework is developed to tackle this problem. The first layer establishes an approximate cost evaluation model for orbital plane visitation sequences, utilizing a genetic algorithm to identify the optimal sequence from a vast array of candidate planes, thereby maximizing inspection targets while minimizing fuel consumption. The second layer constructs a mixed-integer programming model to locally refine the rendezvous epochs and orbital parameters of each inspection orbit to reduce the total velocity increment. The third layer accurately computes the optimal impulsive maneuvers and trajectories between inspection orbits. In contrast to traditional low-Earth orbit rendezvous optimization frameworks, the proposed framework fully leverages the adjustable freedom in inclination and right ascension of the ascending node (RAAN) of inspection orbits, significantly reducing the total velocity increment. Simulation results demonstrate that the proposed method can effectively address the trajectory optimization problem associated with constellation inspection for tens of thousands of satellites.
\end{abstract}



\begin{keyword}
Constellations Inspection \sep Maneuver-Free Inspection Orbit \sep Multi-flyby Trajectory Optimization  \sep Orbital-Plane-Based Strategy
\end{keyword}

\end{frontmatter}

\section{Introduction}
\label{sec1}

The rapid expansion of large satellite constellations by multiple countries is expected to increase the number of low-Earth-orbit (LEO) satellites to tens of thousands in the near future \cite{1,2}. This increased satellite density presents considerable challenges for the management of the space traffic environment. In the event of a satellite malfunction that renders it uncontrollable, the risk of collision incidents escalates, potentially triggering a cascading effect known as the Kessler Syndrome \cite{3,4}. Therefore, it is essential to monitor satellite conditions in a timely manner. However, the distinctive characteristics of the space environment complicate the comprehensive evaluation of satellite status when relying solely on telemetry data, particularly in assessing damage to external structures caused by impacts from charged particles or small debris. The development of a cost-effective inspection mission capable of conducting regular external assessments of satellites could significantly enhance the operational safety and reliability of these large constellations. 

The inspection methods typically manifest in two forms: flybys and rendezvous \cite{5, ARSHAD2025996}. The rendezvous approach entails the spacecraft achieving coincident position and velocity with the target satellite, making it suitable for detailed observation and subsequent operations such as maintenance and fuel replenishment. However, in an era characterized by a vast number of satellites, the costs associated with such operations are likely to increase substantially. In contrast, the flyby method effectively minimizes fuel expenditures while still satisfying the requirement for photographic evaluations of target satellites. Therefore, this study primarily focuses on the traversal inspection of multiple satellites across various orbital planes utilizing the flyby method. From an economic standpoint, trajectory design for such operational spacecraft is advantageous in maximizing access to numerous satellites. 

Existing researches on multi-target trajectory optimization predominantly center on rendezvous-type in-orbit service \cite{6,7,8} or space debris removal missions \cite{9,10,11,12}. In scenarios involving coplanar traversal rendezvous, Shen \cite{6} posited that the optimal rendezvous order should follow the direction of increasing or decreasing phases. Zhang \cite{7} investigated the in-orbit service path planning for quasi-coplanar satellites with minimal phase and right ascension of ascending node (RAAN) offsets, employing a double-impulse maneuver strategy based on differential orbital elements to approximate required velocity increments. Additional studies \cite{13, 14, 15}  have developed rapid $\Delta v$ approximation techniques for multi-impulse rendezvous in near-circular orbits, thereby enhancing the optimization efficiency of multi-rendezvous problems. Regarding the flyby problem, Yan \cite{16} introduced a rapid method for estimating optimal flyby $\Delta v$, applying it to the trajectory optimization of multi-flyby inspection missions. However, practical experience indicates that the efficiency of such algorithms diminishes exponentially as the number of targets increases, rendering them inadequate for the global optimization of flyby paths for large-scale constellations comprising thousands or even tens of thousands of satellites. 

To reduce the complexities associated with flyby path planning for large-scale satellites across multiple orbital planes, this study examines two critical aspects. First, for satellites within the same orbital plane, their phases are typically uniformly distributed. If the initial semi-major axis of the spacecraft is appropriately differentiated from that of the satellites, such that the phase drift per orbit of the spacecraft matches the phase difference between adjacent satellites, sequential flybys of all satellites within the orbital plane can be achieved without  maneuvers. Second, based on the natural RAAN drift in LEO \cite{17, 18}, if the semi-major axis and inclination of two orbital planes are different, the difference in their RAAN drift rates would result in perodic epochs when their RAAN values become euqal, creating low-$\Delta v$ transfer opportunities between the planes. Therefore, for constellations with varying altitudes and inclinations, an optimal multi-plane sequence can be identified to minimize RAAN differences between adjacent planes. This allows a spacecraft to sequentially rendezvous with their maneuver-free inspection orbits at low total cost. However, there are few studies on this issue so far.  In light of this analysis, this paper proposes a low-cost multi-orbital plane inspection strategy. The operational spacecraft initiates from the inspection orbit of a designated orbital plane and executes sequential flybys of the satellites. Then, maneuvers are implemented to transfer the spacecraft to the inspection orbit of another orbital plane in a short time and flyby the new satellites. Therefore, by carefully optimizing the selection and order of the candidate orbital planes in the target constellations, the efficiency of inspection can be significantly enhanced compared to previous inspection trajectory optimization methods which require performing maneuvers between each satellite flyby.

The major contribution of this study lies in proposing an analytical method for determining maneuver-free inspection orbit parameters enabling satellite flybys within a single orbital plane and designing a global trajectory optimization algorithm for multi-plane inspections under fuel and time constraints. Specifically, an analytical mapping relationship between the inspection orbital elements and the relative position and velocity of the spacecraft with respect to satellites within the designated orbital plane is established based on relative motion equations \cite{19, 20}. Leveraging this mapping, the spacecraft's inspection of an orbital plane can be equivalently modeled as rendezvousing with the corresponding inspection orbit and maintaining for a specified duration. As a result, the path planning problem for inspecting multiple orbital planes is reformulated as a multi-rendezvous problem involving the selection and sequencing of inspection orbits, which is solved by a novel three-layer evolutionary-algorithm-based optimization framework. In contrast to previous optimization frameworks for multi-target rendezvous or flyby missions, the proposed approach allows adjustable inspection orbit parameters with localized flexibility in inclination and RAAN, which effectively reduces the total velocity increment. Simulation results demonstrate that the proposed method can rapidly identify the optimal inspection sequence of candidate orbital planes, particularly in scenarios where the number of orbital planes exceeds hundreds and the number of satellites surpasses tens of thousands. This methodology achieved first place in the 13th China Trajectory Optimization Competition (CTOC13) \cite{21,22}, highlighting its computational superiority.

The subsequent sections are organized as follows: Section 2 describe the problem and the assumptions involved; Section 3 presents the method for calculating the maneuver-free inspection orbit for flybys of targets in a orbital plane; Section 4 outlines a three-layer framework for optimizing the trajectory to maximize the number of inspection satellites; Section 5 presents the simulation results; and Section 6 offers concluding remarks.

\section{Problem Description and Assumptions}
Assuming that the multiple constellations to be inspected consist of $p$ orbital planes, let ${s_j},j = 1,2...p$ denote the number of  satellites in each orbital plane, and ${N_{total}} = \sum\limits_{j = 1}^p {{s_j}} $ denote the total number of satellites. The semi-major axes, inclinations, and RAANs at the initial time of each orbital plane are represented by ${a_j}$, ${i_j}$,  and  ${\Omega _j}({t_0})$, respectively. The satellite phases on the same orbital plane are evenly distributed. The dynamics equation based on the mean orbital elements \cite{18} is expressed by
\begin{equation}
\begin{array}{l}
\frac{{{\rm{d}}a}}{{{\rm{d}}t}} = \frac{{{\rm{d}}e}}{{{\rm{d}}t}} = \frac{{{\rm{d}}i}}{{{\rm{d}}t}} = 0\\
\frac{{{\rm{d}}\Omega }}{{{\rm{d}}t}} =  - \frac{3}{2}J_2^{}{(\frac{{{R_e}}}{p})^2}n\cos i\\
\frac{{{\rm{d}}\omega }}{{{\rm{d}}t}} = \frac{3}{2}J_2^{}{(\frac{{{R_e}}}{p})^2}n(2 - \frac{5}{2}{\sin ^2}i)\\
\frac{{{\rm{d}}M}}{{{\rm{d}}t}} = n + \frac{3}{2}J_2^{}{(\frac{{{R_e}}}{p})^2}n(1 - \frac{3}{2}{\sin ^2}i)\sqrt {1 - {e^2}} 
\end{array}
\end{equation}
where $[a, e, i, \Omega, \omega, M]$ represent the six elements of the spacecraft's orbit,  $n$ represents the mean angular velocity, $p = a(1 - {e^2})$ represents the semi-major axis, ${R_e}$ represents the equatorial radius of the Earth, and $J_2^{}$ represents the second-order harmonic term of the Earth's non-spherical shape.

It should be noted that using the orbit averaging method \cite{18} to predict the orbits in a constellation is more reasonable compared to high-precision dynamic equations based on integration, because the constellation regularly executes maneuvers to eliminates the perturbation effects and maintain its configuration, making the long-term orbital changes more consistent with Eq. (1). Additionally, it can enhance computational efficiency. Considerable researches \cite{18, 23} have been conducted on the transition from average-motion trajectories to high-fieldity trajectories.

The spacecraft performs inspections of the target satellite using a flyby approach, and the magnitudes of relative position and velocity between the spacecraft and target satellite must meet the constraints expressed by
\begin{equation}
\begin{array}{l}
\left\| {{\bf{r}} - {{\bf{r}}_s}} \right\| < \Delta {r_{flyby}}\\
\left\| {{\bf{v}} - {{\bf{v}}_s}} \right\| < \Delta {v_{flyby}}
\end{array}
\end{equation}
where  [${\bf{r}}, {\bf{v}}$] represent the position and velocity of the spacecraft, and ${\bf{r}_s}, {\bf{v}_s}$] represent the position and velocity of the target. 

\begin{figure}[t]
\centering
\includegraphics[scale = 0.8]{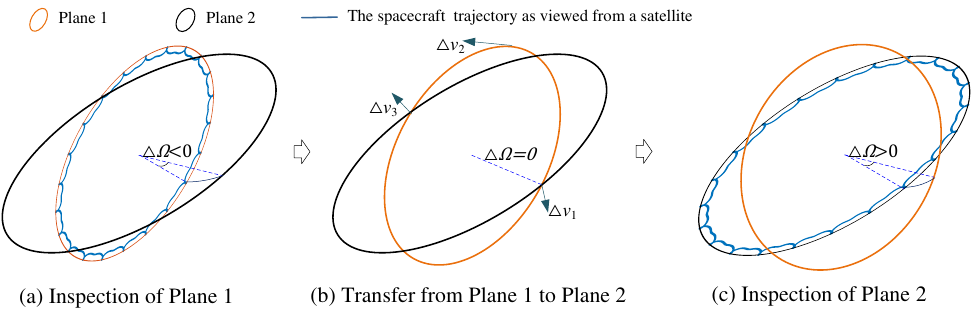}
\caption{Schematic diagram of inspection strategy}\label{fig:fig1}
\end{figure}

The optimization objective is to maximize the number of satellites inspected under fuel and time constraints. The maximum velocity increment for maneuvers is $\Delta {v_{\max }}$, and the mission's start and end times are fixed to $t_0$ and $t_f$. The variables to be determined include the indices of selected orbital planes from the candidate $p$ planes (denoted by $\{ {P_j}\} ,j = 1,2...M$, where $M$ represent the number of selected planes), epochs of arriving each plane (denoted by $\{ {t_j}\} ,j = 1,2...M$), and the impulsive maneuver strategy of the spacecraft (denoted by the epochs $\{ {t_k}\} ,k = 1,2...m$, and three-dimentional velocity increments $\{ \Delta {{\bf{v}}_k}\} ,k = 1,2...m$, where $m$ is the number of maneuvers). The objective function is expressed by
\begin{equation}
J = \sum\limits_{j = 1}^n {{s_{{P_j}}}}
\end{equation}
Based on the analysis presented in the introduction, we implement an inspection strategy that considers the orbital plane as a unit. As illustrated in Fig. \ref{fig:fig1}, after the operational spacecraft reaches a specific orbital plane, it completes the inspection of all targets within the orbital plane by offsetting the semimajor axis to achieve an appropriate phase drift rate. Then, the spacecraft transfers to the next orbital plane that has the minimal RAAN difference relative to the current orbit plane, and continues the inspection. This process is repeated until either fuel consumption or time reaches a predetermined upper limit. Note that when semi-major axis offset is negative, the inspection path is counterclockwise, and the spacecraft's apogee should be near the satellite's orbital altitude; when the semi-major axis difference is positive, the inspection path is clockwise, and the spacecraft's perigee should be near the satellite's orbital altitude. The duration required to complete the inspection of an orbital plane is approximately equal to the product of the number of satellites and the orbital period.
Because the relative distance and velocity during a flyby are also affected by the coupling effects of eccentricity, orbital inclination, RAAN, and the argument of perigee \cite{19, 20}, it is essential to consider these factors comprehensively. To enhance the efficiency of global optimization, the following three assumptions must be clearly defined.

(1) The inclinations of the candidate orbital planes that need to be flyby are relatively similar, which aligns with the actual configurations of existing constellations such as StarLink and OneWeb (e.g., 50, 55, and 60 degrees, among others). Because the velocity increment required for inclination change can not be reduced through long-time drifting caused by the perturbations, and a significant difference in inclination would result in an impractical velocity increment for transfers between inspection orbits. 

(2) To ensure the effectiveness of close-range inspections, the relative distance and velocity during flybys of each satellite are constrained within a limited range. In this condition, the spacecraft must adjust its orbital plane to be in proximity to the target orbital plane through specific maneuvers; however, exact alignment is not required. There exists a degree of flexibility in optimizing the inclination difference and RAAN difference relative to the target orbital plane, which can contribute to fuel efficiency.

(3) The altitude of an inspection orbit must be sufficiently high to prevent atmospheric entry. Therefore, this study only considers inspection orbits characterized by a positive offset in the semimajor axis. Additonally, it is assumed that observations are only performed at the ascending or descending nodes of the satellites, which keeps the perigees of the inspection orbits unchanged to avoid additional $\Delta v$ reqiured for direction change in the arch line. Moreover, this assumption ensures the relative position is only related to $\Delta \Omega $ and the relative velocity is only related to $\Delta i$, which enhances computational efficiency. In this scenario, since the angle between the solar vector and the equatorial plane is always less than 23 degrees, the sunlight conditions during an inspection can be easily assured by making slight adjustments to the perigee altitude of the inspection orbit. This is due to the fact that, when observing from below the satellite's radial position and from above, there will always be one scenario that ensures the camera is in front lighting condition. It should be noted that the observation position can be adjusted to other phase angles, and the method presented in this paper remains applicable (the coupling affects of $\Delta \Omega $ and $\Delta i$ to relative position and velocity are considered). 

Based on the aforementioned assumptions, when the semi-major axis, inclination, and RAAN of an orbital plane, and the satellite number are specified, it is possible to analytically determine the maneuver-free inspection orbit. Consequently, the inspection of an orbital plane can be conceptualized as a rendezvous with the corresponding inspection orbit, followed by a certain duration to complete the flybys of all target satellites. The multi-flyby optimization problem, which involves target selection aimed at maximizing Eq. (3), can subsequently be formulated as the selection and sequencing of inspection orbits from the candidate planes of the satellite constellations. This problem bears similarities to the multi-rendezvous optimization problems discussed in \cite{11,12} and can be addressed with relative ease. The framework for solving this problem will be detailed in the subsequent sections.

\section{Analytical Calculation of the Maneuver-Free Inspection Orbit}
This section delineates an analytical approach for the design of maneuver-free orbits intended for the flyby of satellites within a single orbital plane. The inputs includes the start time of the inspection ${t_0}$, the semi-major axis $a_0$, inclination $i_0$, and RAAN ${\Omega _0}$ of the orbital plane, as well as the number of satellites $N$. The output comprises the orbital elements of the inspection orbit represented by $\sigma  = [a,e,i,\Omega ,\omega ,M]$.

A satellite in the orbital plane should be specified as the initial target to flyby. Note that the argument of latitude of this target at ${t_0}$ may not be zero, a wait time is required to ensure that the initial target is at the nodes (assumption (3) of Section 2) to execute the first flyby. The wait time $\Delta {t_0}$ is expressed by
\begin{equation}
\Delta {t_0} = \frac{{2\pi  - {u_0}}}{{{n_0}}}.
\end{equation}
where ${u_0}$ is the argument of latitude of the initial target. Then, $t_0 + \Delta {t_0}$ is used as the actual start time of the inspection. 

As the orbital elements of the spacecraft and the target satellite are relatively similar, and the satellites' orbits are near-circular, the relative position of the spacecraft can be expressed using differential orbital elements in relation to the satellites. The dynamics of the relative position (denoted by $\delta {r_t},\delta {r_n},\delta {r_r}$) in the RTN coordinate system \cite{19, 20, 24} is
\begin{equation}
\begin{array}{l}
\delta {r_t} = {a_0}(\Delta {u_0} + \frac{{ - 3\Delta a}}{{2{a_0}}}{n_0}\Delta t + \Delta \Omega \cos i) + 2{a_0}\Delta e\sin (u - {u_e})\\
\delta {r_n} = \Delta i\sin u - \Delta \Omega \sin i\cos u\\
\delta {r_r} = \Delta a - {a_0}\Delta e\cos (u - {u_e})
\end{array}
\end{equation}
where, $n_0$ is the mean angular velocity of the satellite, $\Delta a$, $\Delta i$, $\Delta e$, $u_e$, $\Delta \Omega$, and $\Delta u_0$ are the difference in semi-major axis, inclination, eccentricity, argument of perigee, RAAN, and argument of latitude at $t_0$, respectively. $\Delta t$ is the time relative to $t_0$, and $u$ is the argument of lattitude of the satellite. $\Delta a$, $\Delta i$, $\Delta e$, $u_e$, $\Delta \Omega$, and $\Delta u_0$ are calculated by
\begin{equation}
\begin{array}{*{20}{l}}
{\Delta a = a - {a_s}}\\
{\Delta {e_y} = e\sin (\omega ) - {e_t}\sin ({\omega _s})}\\
{\Delta {e_x} = e\cos (\omega ) - {e_t}\cos ({\omega _s})}\\
{\Delta u = \omega  + M - ({\omega _s} + {M_s})}\\
{\Delta i = i - {i_s}}\\
{\Delta \Omega  = \Omega  - {\Omega _s}}
\end{array}
\end{equation}
where $a, e, i, \Omega, \omega, M$ are the classical orbital elements of the spacecraft, and the subscript 's' represents the orbital elements of the satellite. Once the differential orbital parameters are determined, the inspection orbit parameters and the relative motion of the spacecraft can be derived. The process of determining the inspection orbit is detailed as follows. 

First, to ensure that all targets are sequentially approached during the flyby, the argument of latitude of the operational spacecraft should lag behind the initial target by $\frac{{2\pi }}{N}$ per orbit. Then, the orbital period of the spacecraft is $\frac{{(N + 1)}}{N}{T_0}$ (where $T_0$ is the orbital peroid of target satellites), and the duration after the flyby of the final target satellite is:
\begin{equation}
\Delta {t_{stay}} = \frac{{(N - 1)(N + 1)}}{N}{T_0},
\end{equation}
Additionally, $\Delta a$ is detemined by
\begin{equation}
( - \frac{3}{2}\frac{{\Delta a}}{{{a_0}}}{n_0})(\frac{{2\pi }}{{{n_0}}}) =  - \frac{{2\pi }}{N} \Rightarrow \Delta a = \frac{{2{a_0}}}{{3N}}
\end{equation}
Second, based on the assumption in Section 2, the perigee of the inspection orbit (denoted by ${r_a}$, which determines the relative altitude of flyby) is set to the satellite's altitude plus a small offset $\delta {r_0}$ (e.g., 5 km, set according to safety or other observation reqiurements). Then, $\Delta e$ can be calculated by
\begin{equation}
\begin{array}{l}
{r_a} = {a_0} + \delta {r_0}\\
{r_p} = {a_0} - \delta {r_0} + 2\Delta a\\
\Delta e = ({r_p} - {r_a})/({r_a} + {r_p})
\end{array}
\end{equation}
where ${r_p}$ is the altitude of peregee. 

Third, by differentiating Eq. (5), it is determined that when $u = 0$ (assumption (3) in Section 2), the relative velocity of the spacecraft during the flyby of a target satellite is expressed by
\begin{equation}
\begin{array}{l}
\delta {v_t} = {V_0}(\frac{{ - 3\Delta a}}{{2{a_0}}}{n_0} + 2{a_0}\Delta e)\\
\delta {v_n} = {V_0}\Delta i\\
\delta {v_r} = 0
\end{array}
\end{equation}
where ${V_0} = {a_0}{n_0}$ is the mean velocity of satellites. It can be seen that when the number of satellites is given, ${\Delta a}$ and $\Delta e$ are known, and then $\delta {v_t}$ can be determined. The normal componet $\delta {v_n}$ is proportional to $\Delta i$. Therefore, when $\delta {v_t} > \delta {v_{flyby}}$ holds, the relative velocity constraint in Eq. (2) can never be satisfied and the maneuver-free strategy is infeasible; otherwise, when $\delta {v_t} < \delta {v_{flyby}}$ holds, the permissible range of $\Delta i$ is expressed by
\begin{equation}
\Delta i \in [ - \frac{{\sqrt {\delta {v_{flyby}}^2 - \delta {v_t}} }}{{{V_0}}},\frac{{\sqrt {\delta {v_{flyby}}^2 - \delta {v_t}} }}{{{V_0}}}]
\end{equation}
By defining a normalized coefficient  ${k_i} \in [ - 1,1]$ as a variable to be determined, $\Delta i$ can be expressed by $\Delta i = {k_i}\Delta {i_{\max }}$, where $\Delta {i_{\max }} = \frac{{\sqrt {\delta {v_{flyby}}^2 - \delta {v_t}} }}{{{V_0}}}$. 

Fourth, by substituting $\Delta a$ and $\Delta e$ back into Eq. (5) and setting $u = 0$ (assuming (3) in Section 2), the three components of relative position at the initial time when the spacecraft approaches the first satellite are $\delta {r_0}$, $\Delta {u_0} + \Delta \Omega \cos i$, and ${a_0}\Delta \Omega \sin i$. For the normal component, due to the influence of $J_2$ perturbation, RAAN and the argument of perigee both experience a certain drift. Therefore, when the spacecraft approaches the second through the $N^{th}$ satellite, variations in $\Delta \Omega $ and ${u_e}$ in Eq. (5) must be considered. The drift rate of $\Delta \Omega $ is the difference in the RAAN drift rates between the spacecraft and the satellite, calculated by
\begin{equation}
\frac{{{\rm{d}}\Delta \Omega }}{{{\rm{d}}t}} = ( - 3.5\frac{{\Delta a}}{{{a_0}}} - \tan i\Delta i)\frac{{{\rm{d}}\Omega }}{{{\rm{d}}t}}
\end{equation}
It can be seen that when the spacecraft arrives at the final satellite, the change in $\Delta \Omega $ is
\begin{equation}
\Delta {\Omega _d} = \Delta {t_{stay}}( - 3.5\frac{{\Delta a}}{{{a_0}}} - \tan i\Delta i)\frac{{{\rm{d}}\Omega }}{{{\rm{d}}t}}
\end{equation}
Thus, the change in $\delta {r_n}$ is ${a_0}\Delta {\Omega _d}\sin i$. Because the relative argument of lattitude of each flyby remains zero ($\delta {r_t}$ = 0 ) and $\delta {r_r} = {\delta r_0}$, the maximum $\delta {r_n}$ that satisfies the constraint of Eq. (2) is $\sqrt {\delta r_{flyby}^2 - \delta r_0^2}$. Therefore, when $\left| {{a_0}\Delta {\Omega _d}\sin i} \right| = 2\sqrt {\delta r_{flyby}^2 - \delta r_0^2} $ holds, $\Delta {\Omega _0}$ corresponding to the first flyby can be set to $\frac{{ - \Delta {\Omega _d}}}{2}$. When $\left| {{a_0}\Delta {\Omega _d}\sin i} \right| > 2\sqrt {\delta r_{flyby}^2 - \delta r_0^2} $ holds, Eq. (2) can not be satisfied. Otherwise, when $\left| {{a_0}\Delta {\Omega _d}\sin i} \right| < 2\sqrt {\delta r_{flyby}^2 - \delta r_0^2}$, there exists a certain degree of flexibility for $\Delta {\Omega _0}$, which is expressed by
\begin{equation}
\Delta {\Omega _0} \in [ - \frac{{\sqrt {\delta r_{flyby}^2 - \delta r_0^2} }}{{{a_0}\sin i}} + \left| {\frac{{\Delta {\Omega _d}}}{2}} \right|,\frac{{\sqrt {\delta r_{flyby}^2 - \delta r_0^2} }}{{{a_0}\sin i}} - \left| {\frac{{\Delta {\Omega _d}}}{2}} \right|]
\end{equation}
Thus, by introducing another normalized coefficient  ${k_\Omega} \in [ - 1,1]$ as a variable to be determined, $\Delta {\Omega _0}$ can be expressed by ${k_\Omega }(\frac{{\sqrt {\delta r_{flyby}^2 - \delta r_0^2} }}{{{a_0}\sin i}} - \left| {\frac{{\Delta {\Omega _d}}}{2}} \right|)$. 

Fifth, according to the drift rate of the argument of perigee \cite{18}, the change in $\omega$ during the inspection is 
\begin{equation}
\Delta {\omega _d} = \Delta {t_{stay}}( - 3.5\frac{{\Delta a}}{{{a_0}}} - 5\sin i\cos i\Delta i)\dot \omega
\end{equation}
To ensure that ${u_e}$ is close to zero, the initial $\omega$ of the spacecraft at $t_0$ should be offsetted to $\frac{{ - \Delta {\omega _d}}}{2}$. Treating $\Delta {\omega _d}$ as a small angle and substituting it into Eq. (5), $\delta r_r$ approximately remains zero, but $\delta r_t$ must be modified to
\begin{equation}
\begin{array}{l}
\delta {r_t}({t_0}) = {a_0}(\Delta {u_0} - \frac{{\Delta {\Omega _d}}}{2}\cos i) - {a_0}\Delta e\Delta {\omega _d}\\
\delta {r_t}({t_0} + \Delta {t_{stay}}) = {a_0}(\Delta {u_0} + \frac{{ - 3\Delta a}}{{2{a_0}}}{n_0}\Delta {t_{stay}} + \frac{{\Delta {\Omega _d}}}{2}\cos i) + {a_0}\Delta e\Delta {\omega _d}
\end{array}
\end{equation}
where $\delta {r_t}({t_0})$ and $\delta {r_t}({t_0} + \Delta {t_{stay}})$ represent the tangential relative positions at the start and end epochs of the inspection. Therefore, to ensure $\delta {r_t}({t_0})$ and $\delta {r_t}({t_0} + \Delta {t_{stay}})$ remain zero, the initial difference in argument of latittude ($\Delta {u_0}$) should be set to
\begin{equation}
\Delta {u_0} = \Delta e\Delta {\omega _d} + \frac{{\Delta {\Omega _d}}}{2}\cos i
\end{equation}
and the semimajor axis offset should also be corrected by $\Delta {a_c}$, which is calculated by
\begin{equation}\frac{{ - 3\Delta {a_c}}}{2}{n_0}\Delta {t_{stay}} = \delta {r_t}({t_0} + \Delta {t_{stay}}) \Rightarrow \Delta {a_c} = \frac{{2\delta {r_t}({t_0} + \Delta {t_{stay}})}}{{ - 3{n_0}\Delta {t_{stay}}}}
\end{equation}
In summary, the analytical method for calculating the inspection orbit elements is outlined in Algorithm 1. Due to the presence of $k_i$ and $k_\Omega$, the inspection orbit is not uniquely determined by the orbital plane to flyby, thereby introducing two degrees of freedom. This constitutes the primary difference between the multi-plane inspection orbit optimization problem addressed in this study and traditional multi-satellite rendezvous problems. 
\begin{algorithm}[H]
\caption{Calculation of Inspection Orbital Elements}
\label{alg:1}
\begin{algorithmic}[1] 
\Require $a_0$, $i_0$, $\Omega_0$ of the orbital plane, satellite count $N$, $u_0$ of the initial satellite, $k_i$, and $k_\Omega$
\Ensure $\Delta a$, $\Delta i$, $\Delta e$, $u_e$, $\Delta \Omega$, and $\Delta u_0$ relative to the initial satellite, time reqiurement $\Delta {t_{stay}}$
\State Modify $t_0$ by Eq. (4)
\State Calculate $\Delta {t_{stay}}$ by Eq. (7) 
\State Calculate $\Delta a$ and $\Delta e$ by Eq. (8)
\State Set  $\Delta i = {k_i}\frac{{\sqrt {\delta {v_{flyby}}^2 - \delta {v_t}} }}{{{V_0}}}$
\State Set $\Delta {\Omega _0} = {k_\Omega }(\frac{{\sqrt {\delta r_{flyby}^2 - \delta r_0^2} }}{{{a_0}\sin i}} - \left| {\frac{{\Delta {\Omega _d}}}{2}} \right|)$
\State Calculate $\Delta {\omega _d}$ by Eq. (12) and set  $\Delta {\omega _0} =  - \Delta {\omega _d}/2$
\State Calculate $\Delta {u_0}$  by Eq. (17)
\State Correct $\Delta a$ by Eq. (18)
\end{algorithmic}
\end{algorithm}

\section{Global Optimization Algorithm for Multi-Plane Flyby Inspection Trajectory}
The optimization framework presented in this section is structured in three layers, as illustrated in Fig. 2. The first layer employs an integer-encoded genetic algorithm (GA) for global search of orbital plane sequences, where the flyby initiation time for each orbital plane and the corresponding parameters ($k_i$, and $k_\Omega$) in Algorithm 1 are adaptively determined using an approximation strategy rather than being treated as unknown variables to enhance efficiency. The second layer maintains the selected orbital planes and applies a mixed-integer-encoded differential evolution (DE) algorithm to re-optimize both the visitation sequence of the selected orbital planes and the flyby initiation times while simultaneously incorporating each inspection orbit's $k_i$, and $k_\Omega$ as unknown variables, with the orbjective of minimizing the total velocity increment. Finally, the third layer fixes each inspection orbit and rendezvous timing to solve for the spacecraft's impulsive transfer trajectory by directly implementing the four-impulse optimization algorithm for near-circular low-Earth-orbit rendezvous in \cite{23}. This study primarily focuses on the algorithmic design of the first two layers, particularly the encoding schemes of unknown variables and expression of objective functions, while refraining from modifying the evolutionary algorithms themselves.
\begin{figure}[htb!]
\centering
\includegraphics[scale = 0.9]{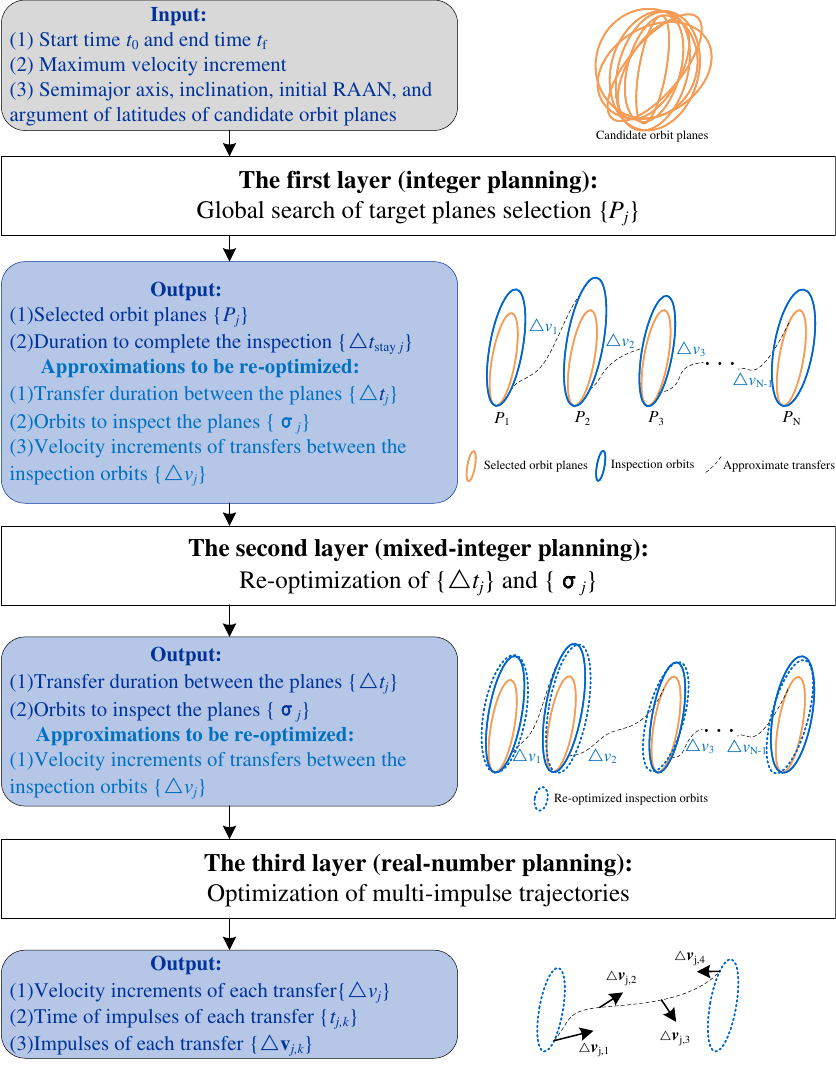}
\caption{Three-layer optimization framework}\label{fig:fig2}
\end{figure}

\subsection{Global search for orbital plane selection and sequencing}
The indices of the candidate orbital planes are used as the encoded vector $\bf{x}$ in GA. Since the optimal sequence length is not predetermined, we define a maximum allowable sequence length, $L$, as the fixed encoding dimension. The actual sequence length is obtained by truncating the trailing orbital planes that either surpass time constraints or violate velocity increment limits. Thus, $\bf{x}$ is expressed by
\begin{equation}
{\bf{x}} = \{ {P_j}\} ,j = 1,2...L,{P_j} \in [1,{N_{total}}]
\end{equation}
According to Algorithm 1, the parameters of the inspection orbits (denoted by ${{\bf{\sigma }}_j}$) in a sequence can be calculated by inputting the start time of flyby for each orbital plane ($t_j$), along with the coefficients ${k_{i,j}}$ and ${k_{\Omega ,j}}$ (where the subscript '$j$' represents the $j^{th}$ plane).  To avoid optimizing real-valued variables during the sequence search, the inputs for Algorithm 1 in the first layer are simplified by setting ${k_{\Omega ,j}} = 0, j = 1,2...L$ and ${k_{i,1}} = 0$. Then, ${{\bf{\sigma }}_1}$ and $\Delta {t_{stay,1}}$are obtained. When $j > 1$, ${k_{i,j}}$ is determined through the following steps. 

First, calculate $\Delta {i_{\max ,j}}$ according to Eq. (11). Then, determine $\Delta i$ by minimizing the absolute difference in inclination between the previous and current inspection orbits via
\begin{equation}
\Delta i = \left\{ \begin{array}{l}
\Delta {i_{\max }},i + \Delta {i_{\max }} < {i_{j - 1}}\\
 - \Delta {i_{\max }},i - \Delta {i_{\max }} > {i_{j - 1}}\\
i - {i_{j - 1}},{i_{j - 1}} - \Delta {i_{\max }} < i < {i_{j - 1}} + \Delta {i_{\max }}
\end{array} \right.
\end{equation}
${k_{i,j}}$ is thus set to $\Delta i/\Delta {i_{\max }}$. Eq. (23) is a local-optimal strategy that aims to minimize the $\Delta v$ reqiured for outer-plane maneuvers. 
$t_j$ can be approximated as follows. First,  $t_1$ is euqla to $t_0$. Then, let $\Delta {t_j}$ denote the transfer durations from the $j-1^{th}$ inspection orbit to the $j^{th}$ inspection orbit, $t_j$ is given by
\begin{equation}
{t_j} = {t_{j - 1}} + \Delta {t_j} + \Delta {t_{stay,j}}
\end{equation}
$\Delta {t_j}$ is determined via an approximate methodology. As the RAAN drift rates for both previous and current inspection orbits are known, we can identify the time corresponding to the minimum RAAN difference as the optimal moment to rendezvous with the current orbit. This local-optimal strategy includes two steps as follows. 

First, because the optimization objective is to maximize the number of inspected satellites, the transfers between inspection orbits should avoid excessive time consumption. Therefore, the range of $\Delta {t_j}$ is set to
\begin{equation}
\Delta {t_j} \in [\Delta {t_{\min }},\Delta {t_{\max }}]
\end{equation}
where $\Delta {t_{\min }}$ and $\Delta {t_{\max }}$ are hyperparameters that must be predetermined and can be adaptively adjusted based on the actual performance of the optimization process.
Second, calculate the RAAN difference between ${{\bf{\sigma }}_{j - 1}}$ and ${{\bf{\sigma }}_j}$ at ${t_{j - 1}} + \Delta {t_{\min }} + \Delta {t_{stay,j-1}}$ and ${t_{j - 1}} + \Delta {t_{\max }} + \Delta {t_{stay,j-1}}$, respectively (denoted by $\Delta {\Omega _1}$ and $\Delta {\Omega _2}$). As the difference in RAAN drift rate ($\Delta \dot \Omega $) between ${{\bf{\sigma }}_{j - 1}}$ and ${{\bf{\sigma }}_j}$ is fixed, $\Delta {t_j}$ can be determined under two distinct scenarios: when $\Delta {\Omega _1}$ and $\Delta {\Omega _2}$ share the same sign, $\Delta {t_j}$ takes the moment with smaller absolute value between $\Delta {\Omega _1}$ and $\Delta {\Omega _2}$; when $\Delta {\Omega _1}$ and $\Delta {\Omega _2}$ have opposite signs, $\Delta {t_j}$ is the moment that nullifies the RAAN difference, which can be expressed by
\begin{equation}
\Delta {t_j} = \Delta {t_{\min }} + \frac{{\Delta {\Omega _2} - \Delta {\Omega _1}}}{{\Delta \dot \Omega }}
\end{equation}

When the coefficients ${k_{i,j}}$ and ${k_{\Omega ,j}}$, and ${t_j}$ are obtained by Eqs. (23), (24), and (26), they are substituted into Algorithm 1 to calculate the $j^{th}$ inspection orbit ${{\bf{\sigma }}_j}$ in the sequence. It should be noted that, according to Eq. (4), the actual initiation time of each inspection orbit (when satellites reach perigee) varies depending on which satellite within the orbital plane serves as the starting point, consequently causing variations in the output. To identify the optimal starting satellite within the orbital plane, it is essential to compute the inspection orbit for each satellite considered as a potential starting point (denoted by ${{\bf{\sigma }}_{j,k}},k = 1,2...{s_j}$), and then select the orbit requiring minimal velocity increment for the transfer from ${{\bf{\sigma }}_{j-1}}$ as the actual inspection orbit (with the corresponding velocity increment denoted as $\Delta {v_j}$). This process utilizes the semi-analytical method in \cite{14} to efficiently estimate $\Delta v$ of these inter-orbital transfers.

In summary, given any specified sequence of orbital planes, it is possible to initiate from the first orbital plane and subsequently determine the corresponding inspection orbit parameters, the transfer dutaion to rendezvous with the subsequent inspection orbit, and the initial satellite of each orbital plane. The total mission duration is accumulated by Eq. (24), the total velocity increment is calculated via$\sum\limits_{j = 1}^L {\Delta {v_j}} $, and the number of flyby satellites is calculated according to Eq. (3). Since the maximum maneuverability of the spacecraft is $\Delta {v_{\max }}$, and the latest end time is $t_f$, constraint checks need to be performed during the calculation. If at the $(m+1)^{th}$ orbital plane in the sequence, $\sum\limits_{j = 1}^{m + 1} {\Delta {v_j}}  > \Delta {v_{\max }}$ or ${t_{m + 1}} > {t_f}$ holds, the calculation must be terminated, and the actual length of the sequence is set to $m$. The total number of inspected satellites is then outputted as $\sum\limits_{j = 1}^m {{s_j}} $.

To further enhance efficiency, a technique from \cite{17} is employed, which incorporates a weighted expression of the total velocity increment into the objective function presented in Eq. (3). The revised expression is
\begin{equation}
J = \sum\limits_{j = 1}^m {{s_j}}  + (1 - \frac{{\sum\limits_{j = 1}^m {\Delta {v_j}} }}{{\Delta {v_{\max }}}})
\end{equation}
This mechanism is designed to guide the population toward fuel-efficient solutions by adjusting the order of orbital planes through the velocity increment weighting term when the number of orbital planes and flyby satellites become confined within local optima, thereby indirectly creating opportunities to insert additional orbital planes into current sequence. The calculation of the objective function (i.e., fitness in the GA) is outlined in Algorithm 2. 
\begin{algorithm}[htb!]
\caption{Calculation of objective function}
\label{alg:2}
\begin{algorithmic}[1] 
\Require Sequence of orbit planes ${\bf{x}} = \{ {P_j}\} $
\Ensure $J$
\State Set ${k_{\Omega ,j}} = 0,j = 1,2...L$ and ${k_{i,1}} = 0$
\State Calculate ${{\bf{\sigma }}_1}$ and $\Delta {t_{stay,1}}$ by Algorithm 1
\For $j=2:L$
    \State Calculate ${k_{i,j}}$ by Eq. (20)
    \For  $k=1:s_j$ (each satellite in the current orbital plane)
        \State Calculate $\Delta {t_{j,k}}$ by Eq. (23)
        \State Calculate ${t_{j,k}}$ by Eq. (22)
        \State Calculate ${{\bf{\sigma }}_{j,k}}$ by Algorithm 1
        \State Calculate $\Delta {v_{j,k}}$ of transfer from ${{\bf{\sigma }}_{j - 1}}$ to ${{\bf{\sigma }}_{j,k}}$
    \EndFor
    \State Set $\Delta {v_j}$ equal to the minimum $\Delta {v_{j,k}}$, and set${{\bf{\sigma }}_{j}}$ to corresponding ${{\bf{\sigma }}_{j,k}}$
    \If $\sum\limits_{k = 1}^j {\Delta {v_k}} > \Delta {v_{\max }}$ or ${t_j} > {t_f}$
        \State Set the actual sequence length m equal to $j-1$
        \State break
    \EndIf
    \State Update $J$ by Eq. (24)
\EndFor
\end{algorithmic}
\end{algorithm}
This study adopts the GA as described in \cite{12, 17} without modifications to the evolutionary process. With a population size of $N_{pop}$ and maximum generation number $G_{max}$, the procedure mainly consists of the following steps: FIrst, randomly generate $N_{pop}$ orbital plane sequences; Second, use Algorithm 2 to calculate the fitness of each individual within the population; Third, apply the selection, crossover, and mutation operators outlined in \cite{17} for population evolution, where the selection operator employs roulette wheel selection to probabilistically eliminate low-fitness individuals, the crossover operator randomly exchanges partial orbital plane sequences between two individuals, and the mutation operator randomly replaces specific orbital planes with alternatives from the candidate set; The second and third steps are repeated until the generation number reaches $G_{max}$. Finally, output the optimal sequence of orbital planes that maximizes the objective function along with the approximately determined inspection orbit parameters (which serve as initial values for the second-layer planning). 

\subsection{Sequence Re-Optimization Algorithm}
The global search algorithm presented in Section 4.1 enhances efficiency by fixing ${k_{\Omega ,j}} = 0$ and utilizing approximation strategies to compute ${k_{i ,j}}$ and $t_j$ of each plane in a given sequence. Once the optimal sequence of orbital planes to flyby has been obtained,  this section ultiliz a real-number encoding DE algorithm \cite{12, 25} to re-optimize ${k_{\Omega ,j}}$, ${k_{i ,j}}$ and $t_j$ in the sequence and adjust the order of selected orbital planes (i.e., inspection orbits) to minimize the total velocity increment. 

Assuming that the length of the obtained orbital plane sequence is $m$, a real-number vector ${x_j} (j=1, ... 4m)$ is used as the normalized undetermined variables. The sorted indices of $\{ {x_j}\} ,j \in [1,m]$ represent the order of rendezvousing the inspection orbits, following the approach presented in \cite{12}. Meanwhile, $\{ {x_j}\} ,j \in [m + 1,2m]$ represent ${k_{\Omega, j} }$, $\{ {x_j}\} ,j \in [2m + 1,3m]$ represent ${k_{i, j}}$, and $\{ {x_j}\} ,j \in [3m + 1,4m]$ represent $\Delta t_j$, as expressed by
\begin{equation}
\begin{array}{l}
\Delta {\Omega _j} = {x_{m + j}}\Delta \Omega {}_{max}\\
\Delta {i_j} = {x_{2m + j}}\Delta i{}_{max}\\
\Delta {t_j} = {x_{3m + j}}(\Delta t{}_{max} - \Delta t{}_{\min }) + \Delta t{}_{\min }
\end{array}
\end{equation}
The objective function in the re-optimization model is calculated as follows. For any given ${x_j}$, the order of the orbital planes can first be obtained by sorting $\{ {x_j}\} ,j \in [1,m]$. Then, the inputs for Algorithm 1 can be derived using Eq. (28) to compute the inspection orbital parameters (${{\bf{\sigma }}_j}$) in the sequence, along with the rendezvous epochs of each inspection orbit (denoted by $t_j$). The $\Delta v$ estimation method in \cite{14} is then utilized to calculate the velocity increment for each transfer (denoted by $\Delta {v_j}$), and the total velocity increment is expressed by
\begin{equation}
J = \sum\limits_{j = 1}^N {\Delta {v_j}} 
\end{equation}
The basic procedure of DE is outlined as follows \cite{25}. First, randomly initalize the population. Second, calculate the objective function of each individual. Third, execute the differential operators and update the population. The second and third steps are iteratively repeated until the maximum generation number is attained, at which point the optimal solution is presented.

Note that in the second-layer optimization, the velocity increment for transfers between adjacent inspection orbits is calculated approximately. The third-layer optimization needs to utilize the multi-impulsive trajectory optimization algorithm as described in \cite{23} by fixing the inspection orbits and the rendezvous epochs to obtain the actual transfer trajectory and maneuver strategy between inspection orbits. The third-layer optimization is not the focus of this study and will not be discussed in further detail.

\section{Simulation Results}
The orbital plane configuration provided by the CTOC13 \cite{21} is used as the simulation scenario for validation. The given dataset consists of 20 constellations, totaling 1,117 orbital planes and 30,188 satellites. However, only 9 constellations with similar inclinations are considered here according to the assumption (1) in Section 2, as detailed in Table 1. Within each constellation, the initial RAAN is uniformly distributed across different orbital planes, and the satellite phases within each orbital plane are also uniformly distributed. The mission duration is constrained to 90 days, with a maximum total velocity increment of 3,000 m/s. Additionally, the maximum allowable relative distance for each flyby is 50 km, and the maximum relative velocity is 150 m/s. This section first validates the inspection orbit calculation method outlined in Section 3. Then, the algorithm presented in Section 4 is tested to identify the optimal single-spacecraft sequence that maximizes the number of inspection targets. Finally, by concatenating multiple sequences involving different target satellites, a collaborative inspection orbit for six spacecraft is achieved. This result won the championship in the CTOC13, thereby demonstrating the effectiveness of the proposed algorithm.
\begin{table}[tbh!]
\centering
\begin{tabular}{p{0.9cm}p{1.3cm}p{1.4cm}p{2.0cm}p{1.6cm}p{1.6cm}p{2.5cm}}
\hline
Index	&Satellite number &Orbital plane number &Satallite number in each plane &Semimajor axis (km) &Inclination (deg) &Inital RAAN of the first plane (deg) \\
\hline
1	&1584	&72	&22	&550.00 	&53.00 	&0.00 \\
4	&1584	&72	&22	&540.00 	&53.20 	&2.50 \\
10	&3600	&60	&60	&508.00 	&60.00 	&4.00 \\
12	&1260	&36	&35	&485.00 	&55.00 	&1.00 \\
13	&900         &30	&30	&600.00 	&55.00 	&2.00 \\
16	&1800	&36	&50	&475.00 	&55.00 	&1.50 \\
17	&1200	&24	&50	&540.00 	&60.00 	&2.50 \\
18	&1200	&24	&50	&550.00 	&60.00 	&3.50 \\
19	&1792	&56	&32	&700.00 	&55.00 	&0.25 \\
\hline
\end{tabular}
\caption{Parameters of constellations to be inspected}\label{tab1}
\end{table}

\subsection{Verification of inspection orbit calculation}
In the subsequent simulation, we use $k-j$ to designate the $k^{th}$ orbital plane in the $j^{th}$ constellation. Taking the 1-1 orbital plane as an example (where the satellite number is 22), by substituting the orbital parameters into Algorithm 1 and setting $\delta r_0$ = 5 km, $k_\Omega = 0$, $k_i = 0$, we can obtain the inspection orbital elements starting from the first satellite at $t_0$ = 0 as $\bf{\sigma }$ = [7136.437 km, 0.0285858, 0.9250245 rad, -0.0055751 rad, -0.0346827 rad, 0.0358305 rad], with a total inspection duration of 1.457 days. The inspection trajectory requires a semi-major axis offset of 210 km, an initial RAAN offset of 0.0056, and an argument of perigee offset of 0.035 relative to the target satellite's orbit.

Then, using the absolute dynamics equations (Eq. (1)) to propagate both the operational spacecraft's and the satellite's orbits, the variations in relative position and relative velocity during the inspection of all 22 satellites at perigee can be obtained, as illustrated in Figure 3. The results prove that the analytical method proposed in Section 3 is capable of facilitating passive flybys of all satellites. Specifically, the radial distance is approximately equal to $\delta r_0$, with deviations within 0.2 km; the along-track distance is approximately zero, with deviations within 0.2 km. These deviations primarily arise from the difference between linear relative motion equations and Eq. (1). The cross-track distance reflects the difference in RAAN between the spacecraft and the satellites, which changes from positive to negative under the influence of $J_2$ perturbation, exhibiting equal magnitudes for the maximum positive and negative distances. This result proves that the RAAN offset strategy can minimize the maximum cross-track distance.
\begin{figure}[hbt!]
\centering
\subfloat[Relative position]{\includegraphics[width=0.45\textwidth,height=0.34\textwidth]{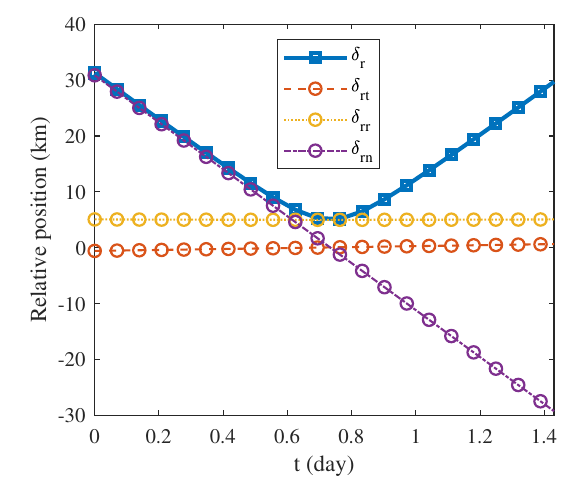}}
\label{subfig3a}
\hfil
\subfloat[Relative velocity]{\includegraphics[width=0.45\textwidth,height=0.34\textwidth]{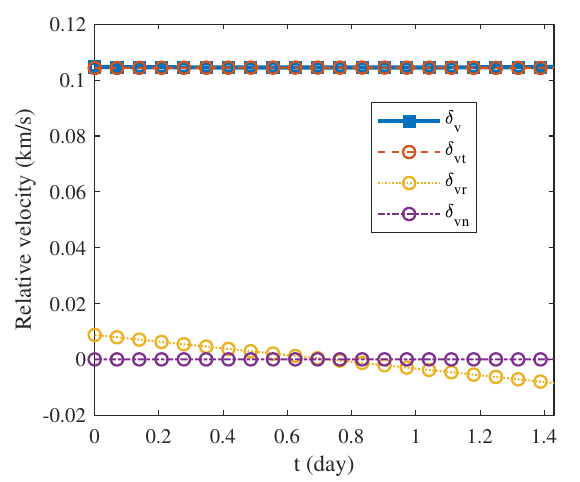}}
\label{subfig3b}
\caption{Variation of relative position and velocity during the inspection of each satellite}
\label{fig:fig3}
\end{figure}
As illustrated in Fig. 3(a), when the permissible maximum relative distance is 50 km, $\frac{{\sqrt {\delta r_{flyby}^2 - \delta r_0^2} }}{{{a_0}\sin i}} - \left| {\frac{{\Delta {\Omega _d}}}{2}} \right|$ in Algorithm 1 is approximately 20 km, corresponding to a maximum RAAN offset of 0.0033 rad. Fig. 3(b) indicates that among the three components of the relative velocity, the along-track velocity caused by $\Delta a$ and $\Delta e$ accounts for the majority proportion, while the radial and cross-track velocities are nearly negligible. The magnitude of relative velocity is approximately 105 m/s. Given that the maximum relative velocity is 150 m/s, the maximum inclination offset calculated by Eq. (11) is 0.014 rad. To validate the constraint satisfaction when $k_\Omega$ and $k_i$ reach their max values, we also calculate the relative positions and velocities corresponding to two cases: $k_\Omega = 1, k_i = 0$ and $k_\Omega = 1, k_i = 1$, which are illustrated in Figures 4 and 5. The results indicate that the inspection orbits with RAAN and inclination offsets can achieve passive inspection while satisfying all constraints. A Comparison of Figures 4 and 5 reveals that  the application of an inclination offset alters the RAAN drift rate, thereby accelerating the rate of change of the cross-track distance ($\delta r_n$) and reducing the allowable maximum RAAN offset while maintaining $\delta r_{flyby}$.
\begin{figure}[hbt!]
\centering
\subfloat[Relative position]{\includegraphics[width=0.45\textwidth,height=0.34\textwidth]{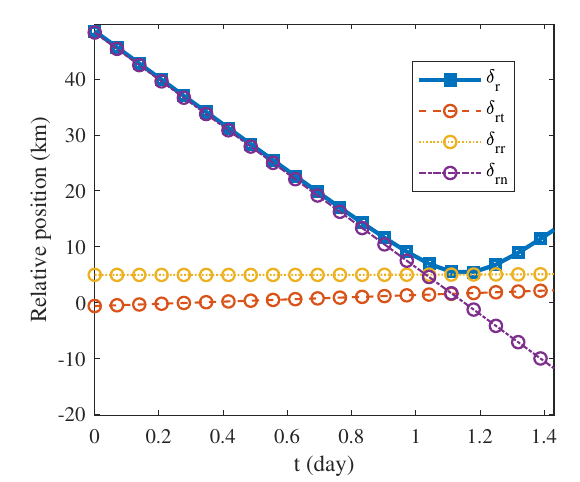}}
\label{subfig4a}
\hfil
\subfloat[Relative velocity]{\includegraphics[width=0.45\textwidth,height=0.34\textwidth]{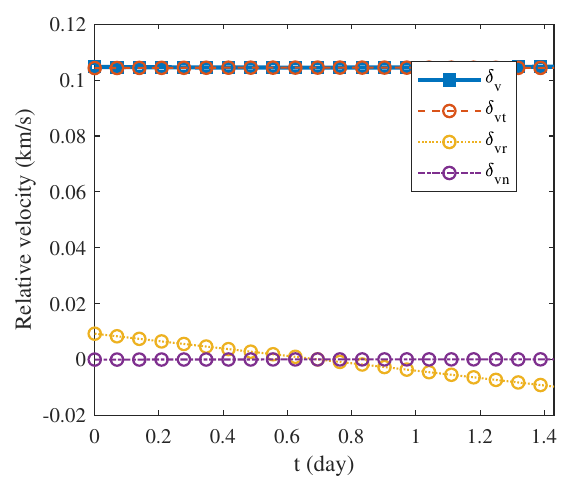}}
\label{subfig4b}
\caption{Variation of relative position and velocity for $k_\Omega = 1, k_i = 0$}
\label{fig:fig4}
\end{figure}
\begin{figure}[hbt!]
\centering
\subfloat[Relative position]{\includegraphics[width=0.45\textwidth,height=0.34\textwidth]{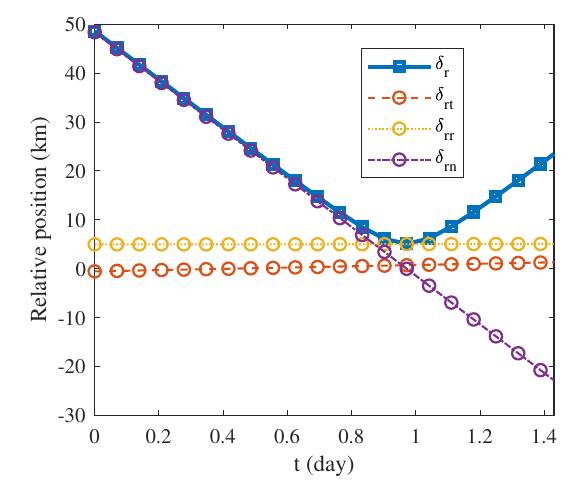}}
\label{subfig5a}
\hfil
\subfloat[Relative velocity]{\includegraphics[width=0.45\textwidth,height=0.34\textwidth]{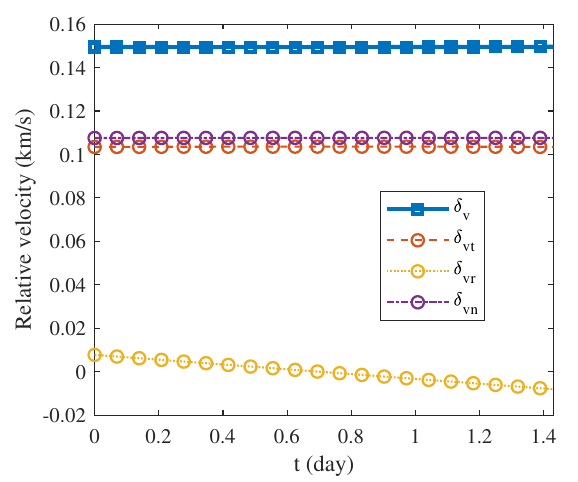}}
\label{subfig5b}
\caption{Variation of relative position and velocity for $k_\Omega = 1, k_i = 1$}
\label{fig:fig5}
\end{figure}
It should be noted that the orbital propagation model employed in this study is an analytical dynamical model that considers only the $J_2$ perturbation (Eq. (1)). If a numerical-integration-based high-fidelity dynamic model \cite{24} (i.e., including perturbations such as lunisolar gravity, atmospheric drag, and higher-order Earth nonspherical gravitational terms) is required, one may first perform trajectory optimization using mean elements, and subsequently obtain the high-fidelity spacecraft trajectory through mean-to-osculating conversion and iterative algorithms \cite{18, 23}.

\subsection{Single-Spacecraft Multi-Plane Sequence Optimization}
This subsection employs the proposed global optimization framework to derive the single-spacecraft multi-orbital-plane sequence that maximizes the number of inspected satellites. The orbital parameters of candidate planes are detailed in Table 1. The constraints are as follows: $t_0 =0$, $t_f$ = 90 days, and $\Delta v_{max}$ = 3 km/s. 

The sequence search algorithm (first-layer of the proposed framework) employs multiple approximations in the computation of inspection orbital parameters and evaluations of the transfer durations and velocity increments. The results typically exhibit approximate 25\% reduction in $\Delta v$ after applying the re-optimization algorithm (second-layer of the proposed framework). In this context, we implement the trick in \cite {12} by relaxing $\Delta v_{max}$ to 3.75 km/s during the sequence search. Additionally, $\Delta t_{min}$ is set to 0.1 days and $\Delta t_{max}$ is set to 4 days.

The hyperparameters of GA are established as follows: population size $N_{pop}$ = 60, individual encoding dimension $L$ = 40 (i.e., maximum sequence length), crossover probability $P_c$ = 0.7, mutation probability $P_m$ = 0.3, and maximum number of generations $G_{max}$ = 6000.  The value range of $x_j$ in each individual is constrained between 1 and 410, corresponding to the number of candidate planes.

The algorithm was benchmarked on a standard laptop (4.8 GHz CPU, 32GB RAM), completing computations in approximately 3,000 seconds. Fig. 6 illustrates the distribution of best solutions derived from 200 repeated runs, which indicates the presence of numerous local optima. Solutions inspecting more than 900 satellites occur with 5\% probability, while solutions involving over 800 satellites are observed in 41\% of the cases. This demonstrates the algorithm's proficiency in reliably identifying near-optimal sequences. The best solution achieves inspection of 32 orbital planes (963 satellites) in 90 days, with an estimated $\Delta v$ of 3673 m/s, as detailed in Table 2. Figures 7 through 9 respectively illustrate the histories of the spacecraft's semi-major axis, inclination, and RAAN over 90 days. The optimal sequence prefers to avoid frequent back-and-forth changes in semi-major axis and inclination and leverage natural RAAN drift rates of neighbouring orbital planes to create transfer opportunities, thereby reducing total $\Delta v$. The mission involves 31 inter-plane transfers, of which 26 transfers do not necessitate a RAAN change (transfer duration is $\Delta t_{min}$), and the remaining 5 require the maximum duration ($\Delta t_{max}$) to reduce $\Delta v$. The average transfer duration is 0.7 days, and the mean velocity increment is 118.5 m/s.
\begin{figure}[htb!]
\centering
\includegraphics[scale = 1]{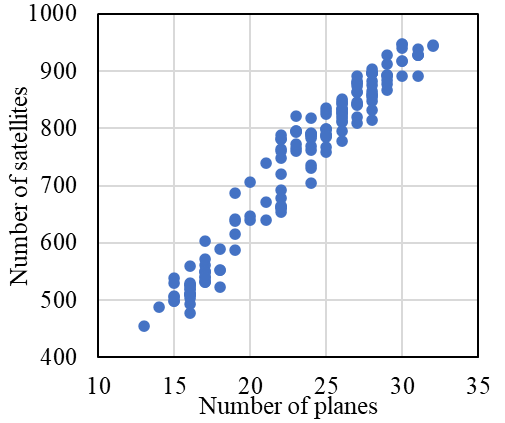}
\caption{Distribution of best solutions from 200 runs}\label{fig:fig6}
\end{figure}

\begin{table}[htb!]
\begin{tabular}{cccc}
\hline
Constellation no. & Orbital plane no. & $\Delta v$ (m/s)           & Satellite number \\ 
\hline
12   & 14    & \textbackslash{} & 35   \\
16   & 14    & 60.1         & 50   \\
4    & 27    & 187.1         & 22   \\
19   & 21    & 102.8         & 32   \\
1    & 28    & 78.0        & 22   \\
4    & 28    & 176.8         & 22   \\
13   & 12    & 57.1         & 30   \\
1    & 29    & 76.5        & 22   \\
4    & 29    & 175.2         & 22   \\
19   & 22    & 150.9         & 32   \\
4    & 31    & 122.6         & 22   \\
16   & 16    & 144.0         & 50   \\
12   & 16    & 46.8         & 35   \\
1    & 32    & 160.6         & 22   \\
4    & 32    & 170.8         & 22   \\
13   & 13    & 85.7         & 30   \\
4    & 33    & 66.9         & 22   \\
16   & 17    & 165.2         & 50   \\
12   & 17    & 56.5         & 35   \\
1    & 34    & 107.3         & 22   \\
1    & 35    & 185.2         & 22   \\
4    & 35    & 166.7         & 22   \\
16   & 18    & 104.5         & 50   \\
12   & 18    & 94.9         & 35   \\
4    & 36    & 166.7        & 22   \\
1    & 37    & 192.4         & 22   \\
19   & 23    & 75.6         & 32   \\
4    & 37    & 91.8         & 22   \\
13   & 14    & 43.3         & 30   \\
1    & 38    & 80.3         & 22   \\
16   & 19    & 151.0         & 50   \\
12   & 19    & 128.6         & 35  \\
\hline
\end{tabular}
\caption{Optimal sequence outputed by the first-layer optimization}\label{tab2}
\end{table}

\begin{figure}[htb!]
\centering
\includegraphics[scale = 0.7]{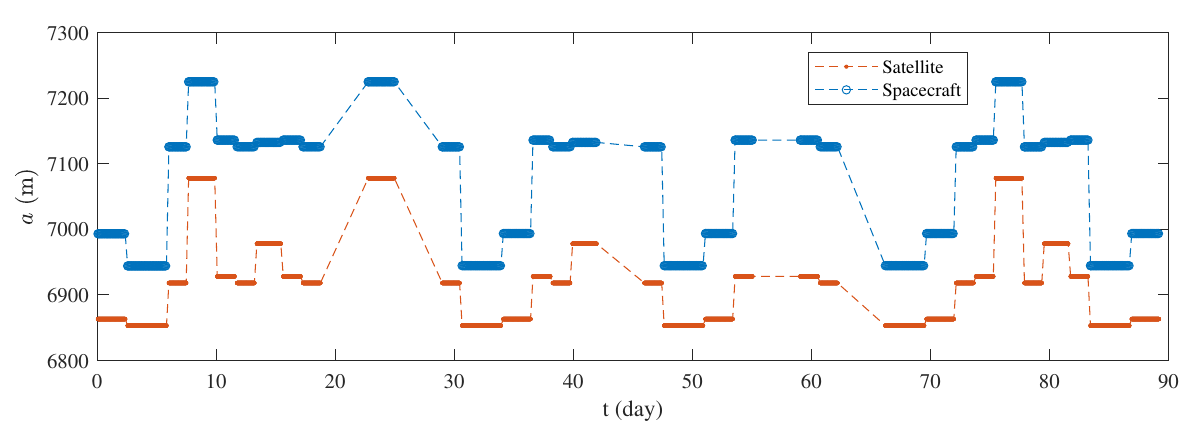}
\caption{History of semimajor aixs of the best sequence}\label{fig:fig7}
\end{figure}
\begin{figure}[htb!]
\centering
\includegraphics[scale = 0.7]{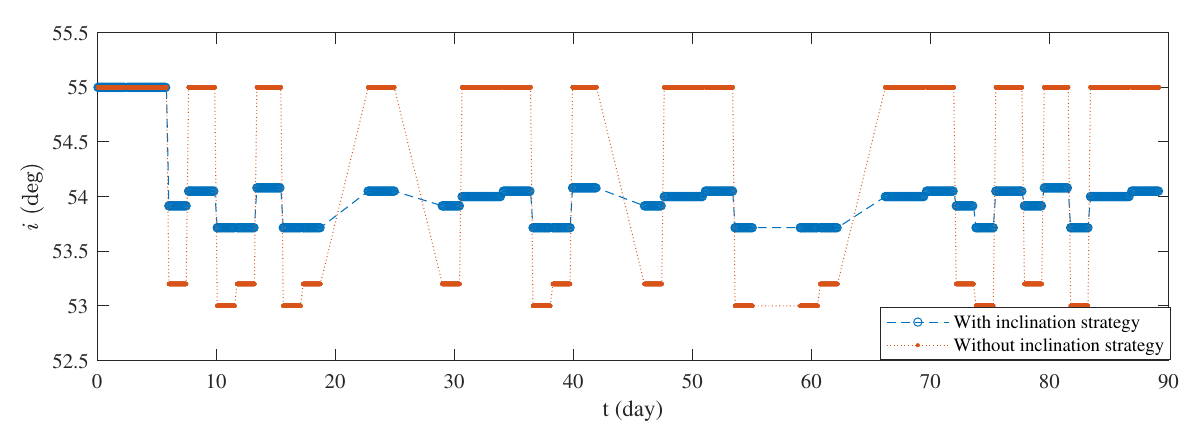}
\caption{History of inclination of the best sequence}\label{fig:fig8}
\end{figure}
\begin{figure}[htb!]
\centering
\includegraphics[scale = 0.7]{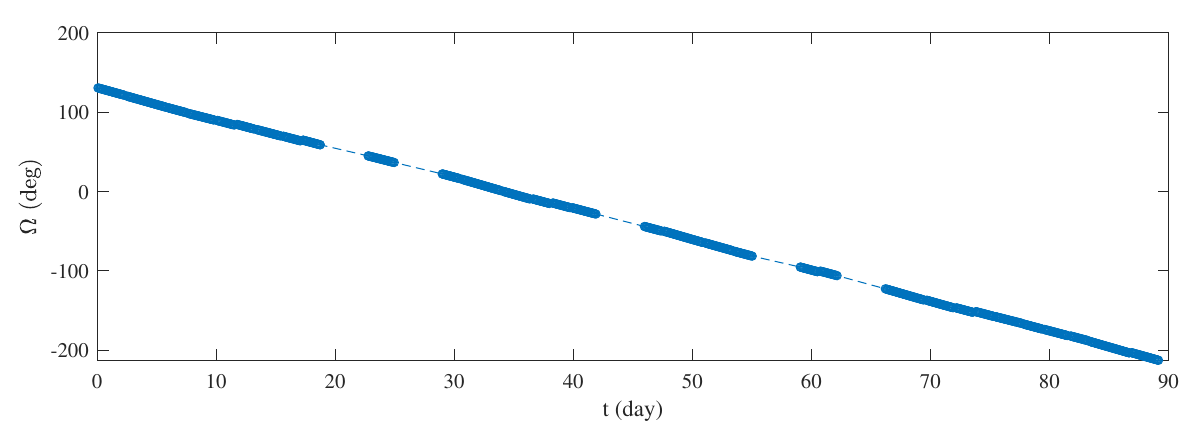}
\caption{History of RAAN of the best sequence}\label{fig:fig9}
\end{figure}
In contrast, when the adaptive inclination offset strategy is not employed (i.e., the spacecraft's inclination must remain aligned with that of each target orbital plane throughout the inspection), the total velocity increment increases to 7.0 km/s. The history of inclination is also illustrated in Fig. 8. These results demonstrate that the proposed inclination offset strategy effectively leverages the maximum relative velocity constraint, thereby avoiding unnecessary orbital inclination adjustments to the spacecraft.

Then, the second-layer optimization algorithm is applied to re-optimize the order of orbital planes, $k_\Omega$ and $k_i$ of each inspection orbit, as well as the flight duration of each transfer.  The result is detailed in Table 3, and the histories of the spacecraft's semi-major axis, inclination, and RAAN of the re-optimized sequence are illustrated in Figures 10 to 12. It is observed that the order remains unchanged; however, refined adjustments of $k_\Omega$, $k_i$, and transfer durations reduced the total velocity increment to 2.95 km/s (a 19.8\% reduction compared to the sequence generated by the first-ayer optimization), demonstrating the importance of the re-optimization model. Additional experiments re-optimizing only rendezvous order and epochs (without re-optimization of $k_\Omega$ and $k_i$) resulted in a total velocity increment of 3.38 km/s, further proving the effectiveness of including inclination and RAAN offsets as variables in the re-optimization process, marking a significant advancement over previous studies.

Note that the velocity increments presented in Table 3 were estimated by the method in \cite{14}. By fixing the inspection orbits and corresponding rendezvous epochs and implementing the multi-impulse trajectory optimization algorithm \cite{23} to obtain the precise trajectory of the operational spacecraft (the third-layer optimization), the actual velocity increment consumption was 2.98 km/s, showing minimal deviation from the estimated results.
\begin{table}[htbp]
\begin{tabular}{p{2.1cm}p{1.8cm}p{1.8cm}p{1.8cm}p{1.8cm}p{1.8cm}}
\hline
Constellation no. & Orbital plane no. & $\Delta v$ (m/s) & Transfer duration (days) & RAAN offset (deg) & Inclination offset (deg) \\
\hline
12   & 14    & \textbackslash{}                                             & \textbackslash{} & 0.00         & 0.00      \\
16   & 14    & 47.4                                                        & 0.11             & 0.00         & -0.29     \\
4    & 27    & 146.6                                                       & 0.64             & 0.09         & 0.72      \\ 
19   & 21    & 55.2                                                        & 0.23             & 0.09         & -0.95     \\
1    & 28    & 66.9                                                        & 0.13             & 0.27         & 0.72      \\
4    & 28    & 127.5                                                       & 0.60             & 0.12         & 0.72      \\
13   & 12    & 48.7                                                        & 0.13             & -0.12        & -0.92     \\
1    & 29    & 87.4                                                        & 0.15             & -0.27        & 0.72      \\
4    & 29    & 84.9                                                        & 1.37             & 0.12         & 0.68      \\
19   & 22    & 171.9                                                       & 5.78             & -0.12        & -0.91     \\
4    & 31    & 63.8                                                        & 2.42             & -0.24        & 0.72      \\
16   & 16    & 101.6                                                       & 0.19             & -0.05        & -0.89     \\
12   & 16    & 28.2                                                        & 0.23             & 0.15         & -0.94     \\
1    & 32    & 121.3                                                       & 0.13             & 0.29         & 0.71      \\
4    & 32    & 78.6                                                        & 1.48             & 0.12         & 0.68      \\
13   & 13    & 159.6                                                       & 0.44             & -0.08        & -0.92     \\
4    & 33    & 124.5                                                       & 0.93             & -0.27        & 0.72      \\
16   & 17    & 117.8                                                       & 0.14             & -0.06        & -0.77     \\
12   & 17    & 50.0                                                        & 0.20             & -0.30        & -0.95     \\
1    & 34    & 94.8                                                        & 0.11             & -0.29        & 0.72      \\
1    & 35    & 118.5                                                       & 5.25             & 0.12         & 0.72      \\
4    & 35    & 73.1                                                        & 1.44             & 0.12         & 0.67      \\
16   & 18    & 252.7                                                       & 0.12             & -0.12        & -0.52     \\
12   & 18    & 31.6                                                        & 0.12             & 0.26         & -0.62     \\
4    & 36    & 183.6                                                       & 0.11             & -0.22        & 0.71      \\
1    & 37    & 114.8                                                       & 1.23             & 0.12         & 0.71      \\
19   & 23    & 66.8                                                        & 0.13             & -0.12        & -0.95     \\
4    & 37    & 56.5                                                        & 0.11             & -0.27        & 0.72      \\
13   & 14    & 37.5                                                        & 0.48             & 0.12         & -0.92     \\
1    & 38    & 57.4                                                        & 0.46             & 0.05         & 0.72      \\
16   & 19    & 107.1                                                       & 0.14             & -0.07        & -1.00     \\
12   & 19    & 70.4                                                       & 0.16             & 0.35         & -0.95    \\
\hline
\end{tabular}
\caption{Optimal sequence outputed by the second-layer optimization}\label{tab3}
\end{table}

\begin{figure}[htb!]
\centering
\includegraphics[scale = 0.7]{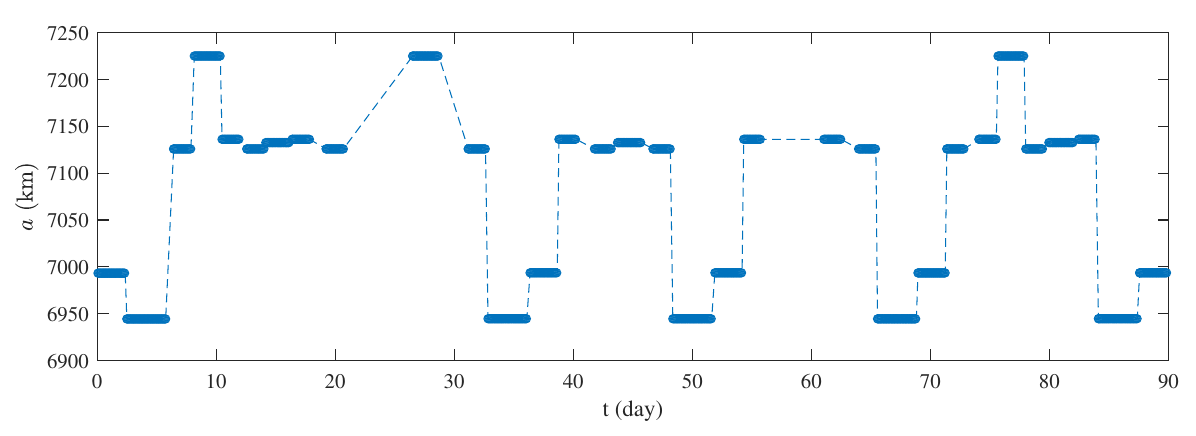}
\caption{History of semimajor aixs of the re-optimized sequence}\label{fig:fig10}
\end{figure}
\begin{figure}[htb!]
\centering
\includegraphics[scale = 0.7]{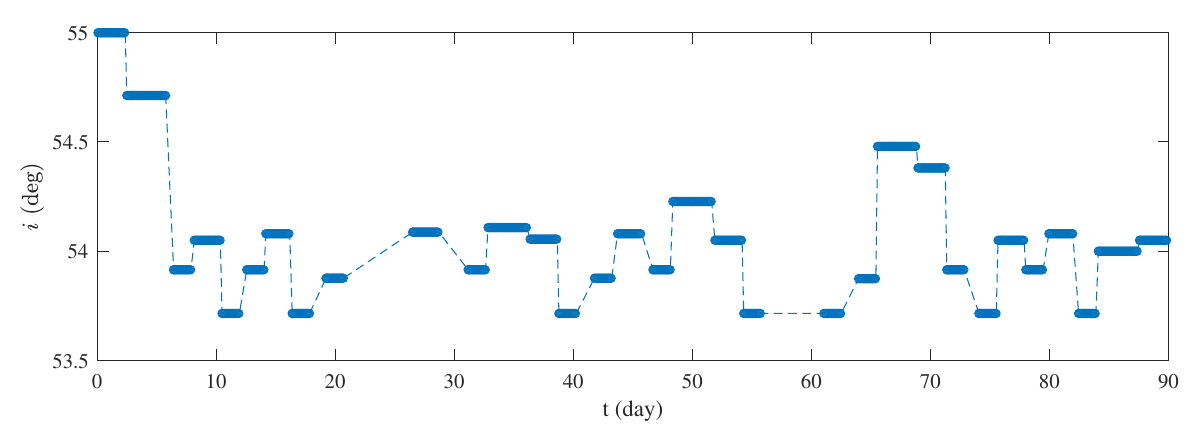}
\caption{History of inclination of the re-optimized sequence}\label{fig:fig11}
\end{figure}
\begin{figure}[htb!]
\centering
\includegraphics[scale = 0.7]{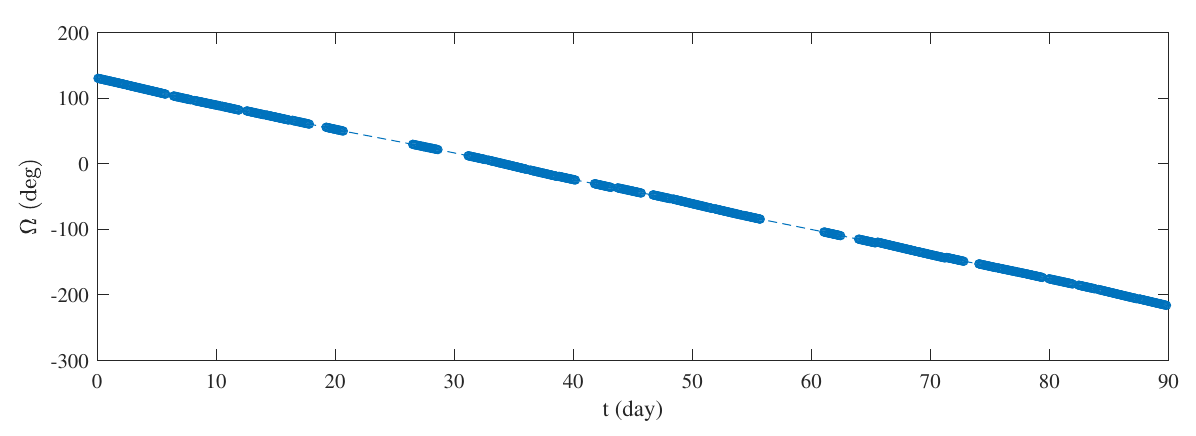}
\caption{History of RAAN of the re-optimized sequence}\label{fig:fig12}
\end{figure}

\subsection{Multi-Spacecraft Solution for CTOC13}
The original problem presented in CTOC13 \cite{21} requires designing optimal trajectories for six operational spacecraft. Each pair of spacecraft must be launched by the same rocket, therefore necessitating that they share the same initial orbit. This scenario represents a multi-spacecraft cooperative multi-flyby problem aimed at maximizing the total number of inspected satellites. In this section, the six spacecraft are labeled as 1-1, 1-2, 2-1, 2-2, 3-1, and 3-2. The problem is reformulated to involve a greedy search for trajectories of six spacecraft sequentially within the candidate orbital plane set, while ensuring no overlapping target satellite and low-$\Delta v$ transfers between the initial orbits of each paired sequences. 

The specific search procedure is outlined as follows. First, use the first-layer optimization algorithm in Section 4 to obtain the inspection sequence of spacecraft 1-1. Second, add the selected orbital planes to a tabu list, and then search for the optimal sequence of spacecraft 1-2 while incorporating the $\Delta v$ of transfer from the fisrt inspection orbit of spacecraft 1-1 to the fisrt orbit of spacecraft 1-2 into the total velocity increment. Third, update the tabu list and repeat the first and second steps twice to generate sequences for spacecraft 2-1, 2-2, 3-1, and 3-2. Fourth, re-optimize the six sequences using the second-layer optimization algorithm in Section 4 to refine the orbital parameters and rendezvous epochs. Finally, each transfer within each sequence should be optimized using the third-layer optimization algorithm to derive the trajectories for each spacecraft.

The solution involving six spacecraft is detailed in Table 4, where the orbital plane index is denoted by the constellation index and the orbital plane index (for instance, 12-14 represents the 14th orbital plane within the 12th constellation). The solution achieves inspections of 143 orbital planes covering 5,516 satellites, and won the CTOC13 championship \cite{22}, thereby demonstrating the effectiveness of the proposed algorithm.

\begin{table}[htbp]
\begin{tabular}{p{2cm}p{5cm}p{2cm}p{2cm}}
\hline
Spacecraft no. & Sequence of orbital planes & Satellite number & Orbital plane number \\ 
\hline
1-1 & 12-14, 16-14, 4-27, 19-21, 1-28, 4-28, 13-12,   1-29, 4-29, 19-22, 4-31, 16-16, 12-16, 1-32, 4-32, 13-13, 4-33, 16-17, 12-17,   1-34, 1-35, 4-35, 16-18, 12-18, 4-36, 1-37, 19-23, 4-37, 13-14, 1-38, 16-19,   12-19 & 963  & 32    \\
1-2 & 10-22, 10-21, 17-9, 10-20, 10-19, 18-8, 17-8,   10-18, 10-17, 18-7, 17-7, 10-16, 10-15, 18-6, 17-6, 10-14                                                                                                            & 890  & 17    \\
2-1 & 12-26, 16-26, 19-40, 13-22, 12-27, 16-27,   4-53, 1-54, 4-54, 1-55, 4-55, 16-28, 12-28, 1-56, 4-56, 13-23, 19-41, 4-57,   16-29, 12-29, 4-58, 1-59, 4-59, 16-30, 12-30, 4-60, 4-61, 13-24, 16-31, 12-31              & 950  & 30    \\
2-2 & 10-41, 18-17, 17-17, 10-40, 10-39, 18-16,   17-16, 10-38, 10-37, 18-15, 17-15, 10-36, 10-35, 18-14, 17-14, 10-34                                                                                                     & 880  & 16    \\
3-1 & 4-8, 13-4, 1-9, 12-5, 16-5, 4-9, 4-10, 19-7,   1-11, 16-6, 12-6, 1-12, 1-13, 4-13, 16-7, 12-7, 1-14, 13-5, 4-14, 1-15, 4-15,   16-8, 12-8, 4-16, 1-17, 19-8, 4-17, 16-9, 12-9, 4-18, 13-6, 1-19                      & 953  & 32    \\
3-2 & 10-6, 18-3, 17-3, 10-5, 10-4, 18-2, 17-2,   10-3, 10-2, 18-1, 17-1, 10-1, 10-60, 18-24, 17-24, 10-59                                                                                                                 & 880  & 16   \\
\hline
\end{tabular}
\caption{Multi-spacecraft solution for CTOC13}\label{tab4}
\end{table}

\section{Conclusions}
This paper addresses the multi-flyby inspection trajectory optimization for mega-constellations in low-Earth-orbit by proposing a global orbital optimization framework to maximize the number of flyby satellites under velocity increment and time constraints imposed on the operational spacecraft. By proposing a maneuver-free inspection strategy and an analytical approach for calculating the corresponding inspection orbits that enable flybys of all satellites within a single orbital plane, the initial problem is effectively reformulated into a multi-rendezvous trajectory optimization problem. This reformulation necessitates only the identification of the optimal sequence of inspection orbits and the optimization of the maneuver strategies between them, thereby significantly reducing computational complexity. Consequently, a three-layer evolutionary-algorithm-based optimization framework is proposed to solve this problem. In the first-layer optimization (selection and sequencing of orbital planes), the integer programming model developed through an adaptive method to determine near-optimal transfer epochs and inspection orbit parameters is efficiently solved using a genetic algorithm. The second-layer optimization further optimizes the total velocity increment by refining the rendezvous epochs and inclination and RAAN offsets of the inspection orbits. Simulation results validate the proposed maneuver-free inspection orbit calculation method and demonstrate the effectiveness of the proposed global trajectory optimization framework for multi-orbit-plane inspection. Through a combination of multiple sequences, the proposed method achieved first place in the CTOC13, proving both its algorithmic superiority and the practicality of conducting rapid inspections of thousands of satellites through rational orbital design.

\section*{Acknowledgement}
This study was supported by the National Natural Science Foundation of China (Grants Nos. 12202504, 12222213, and 1240022077).

%
%
%
\bibliographystyle{unsrt}
\bibliography{En0406}
\end{document}